\documentclass{ws-ijmpd}
\usepackage[super,compress]{cite}
\begin{document}

\markboth{Peter A. Hogan, Dirk Puetzfeld}
{Plane fronted electromagnetic waves and an asymptotic limit of Li\'enard--Wiechert fields}

%
\catchline{}{}{}{}{}
%

\title{PLANE FRONTED ELECTROMAGNETIC WAVES AND AN ASYMPTOTIC LIMIT OF LI\'ENARD--WIECHERT FIELDS}

\author{PETER A. HOGAN}

\address{School of Physics, University College Dublin \\
         Belfield, Dublin 4, Ireland\\
         peter.hogan@ucd.ie}

\author{DIRK PUETZFELD}

\address{University of Bremen, Center of Applied Space Technology and Microgravity (ZARM)\\
        28359 Bremen, Germany\\
        and\\
        School of Physics, University College Dublin \\
        Belfield, Dublin 4, Ireland\\
        dirk@puetzfeld.org}

\maketitle

\begin{history}
\received{Day Month Year}
\revised{Day Month Year}
\end{history}

\begin{abstract}
    Colliding or noncolliding plane fronted electromagnetic or gravitational waves are the asymptotic limit of Robinson--Trautman spherical electromagnetic or gravitational waves. Noncolliding plane fronted waves contain no information about their sources whereas colliding waves contain information about possibly the motion of their sources. As a first step to investigate the latter phenomenon we construct an asymptotic limit of Li\'enard--Wiechert electromagnetic fields in the context of Minkowskian space--time. This has the advantage that the source is well known and the calculations can be carried out in full detail. The final result is an algebraically general Maxwell field which consists of colliding plane fronted waves in a subregion of Minkowskian space--time and an interesting byproduct is a novel perspective on a Maxwell field originally discovered by Bateman.
\end{abstract}


\ccode{PACS numbers: 04.20.-q; 04.20.Jb; 04.20.Cv}



\section{Introduction}\label{sec:1}

We recently \cite{Hogan:Puetzfeld:2021:2} presented a scheme implementing an asymptotic plane fronted limit of Robinson--Trautman \cite{Robinson:Trautman:1960,Robinson:Trautman:1962} spherical electromagnetic and gravitational waves. The resulting plane fronted waves can be either waves with noncolliding wave fronts (the so--called \emph{pp}--waves \cite{Ehlers:Kundt:1962,Pirani:1965,Stephani:etal:2003:book}) or waves with colliding wave fronts (the so--called Kundt \cite{Kundt:1961} waves). The noncolliding waves are disengaged from their sources whereas the colliding waves contain information about their sources such as possibly the motion of the sources. The present paper is a first step to investigate in detail this latter phenomenon. For simplicity and to facilitate explicit detail we restrict ourselves here to electromagnetic fields (in the absence of gravitational fields) for which the source is well known. Hence we take as our starting point Li\'enard--Wiechert electromagnetic fields with source world lines in Minkowskian space--time which are completely general in the sense that they can be time--like, light--like or space--like. Unlike the pure radiation fields considered in Ref.\ \refcite{Hogan:Puetzfeld:2021:2} the Li\'enard--Wiechert field of a point charge is predominantly radiative, with spherical wave fronts, only in a region of Minkowskian space--time at large spatial distance from the world line of the charge. Otherwise the electromagnetic field is algebraically general. This pattern is replicated in the asymptotic limit derived here which is algebraically general but there is a region of Minkowskian space--time in which the asymptotic electromagnetic field is predominantly radiative. In addition the construction described here turns out to be a novel perspective on a Maxwell field originally discovered by Bateman \cite{Bateman:1955:1}{}.

The gravitational analogue of the Li\'enard--Wiechert solutions of Maxwell's vacuum field equations are the Robinson--Trautman solutions \cite{Robinson:Trautman:1960,Robinson:Trautman:1962} of Einstein's vacuum field equations. As well as including the Schwarzschild metric these solutions are the only known exact solutions of Einstein's field equations describing vacuum gravitational fields containing gravitational radiation from isolated sources. The so--called C--metric is an important explicit example of the latter solutions and is currently an active area of research (see, for example Refs.\ \refcite{Astorino:2023:1,Nozawa:Torii:2023:1}). If the radiating point--like sources are accelerating then asymptotically the spherical wavefronts will collide \cite{Hogan:Puetzfeld:2022:book}{}. The Li\'enard--Wiechert electromagnetic fields are a particularly surveyable example of this phenomenon. They can be studied without reference to an external field driving the source since equations of motion are an addition to the field equations in Maxwell's electrodynamics. In the gravitational case the equations of motion of the sources are embedded in the field equations and so a complete description of the asymptotic limit of a Robinson--Trautman space--time requires the addition of an external field which will appear in the equations of motion. This complicates the extension of the work described in this paper to the gravitational case by introducing the Bondi--Sachs generalisation of the Robinson--Trautman solutions (for recent work on the Bondi--Sachs \cite{Bondi:etal:1962:1,Sachs:1962:1} space--times in related contexts see Refs.\ \refcite{Saw:2016:1,Bardeen:Buchman:2012:1,Maedler:2013:1}) and thus will necessarily involve approximations.

Our scheme for constructing an asymptotic limit relies initially on the introduction of two real--valued functions and one complex--valued function via the parametric equations of an arbitrary world line in Minkowskian space--time. Since we are constructing a foundation to effectively facilitate a limit which takes us from the null cone histories of spherical wave fronts emitted by an accelerating charge to the null hyperplane histories of colliding (in general) plane fronted electromagnetic waves, the special parametrization of an arbitrary world line designed to achieve this is not trivial and has been developed in Refs.\ \refcite{Synge:1965:book,Hogan:Puetzfeld:2022:book}{}. This is described in detail in section \ref{sec:2}. A form of the Li\'enard--Wiechert fields exploiting the formalism of section \ref{sec:3} is given in section \ref{sec:4} followed in the next section by the asymptotic limit. The paper ends with a brief discussion in section \ref{sec:5}.

\section{Geometrical Preliminaries}\label{sec:2}

Let $X^i=(X, Y, Z, T)$ with $i=1, 2, 3, 4$ be rectangular Cartesian coordinates and time in Minkowskian space--time with line element
\begin{equation}\label{1}
ds^2=(dX)^2+(dY)^2+(dZ)^2-(dT)^2=\eta_{ij}\,dX^i\,dX^j\ .
\end{equation}
Latin indices take values 1, 2, 3, 4. The summation over repeated indices convention applies and $\eta_{ij}=\eta_{ji}={\rm diag}(1, 1, 1, -1)$. Indices will be raised using $\eta^{ij}$, with $\eta^{ij}=\eta^{ji}$ defined by $\eta^{ij}\,\eta_{jk}=\delta^i_k$, and lowered using $\eta_{ij}$. We use units for which the speed of light in a vacuum $c=1$. An arbitrary world line in Minkowskian space--time has parametric equations $X^i=w^i(u)$ with $u$ an arbitrary parameter along it. Null cones with vertices on this world line have equations $u(X, Y, Z, T)={\rm constant}$ with $u(X, Y, Z, T)$ given implicitly by
\begin{equation}\label{2}
\eta_{ij}\,(X^i-w^i(u))(X^j-w^j(u))=0\ .
\end{equation}
Differentiating this partially with respect to $X^k$ and denoting the partial derivative with a comma we arrive at
\begin{equation}\label{3}
u_{,k}=-\frac{\xi_k}{r}\ \ {\rm with}\ \ \xi_k=\eta_{kl}\,\xi^l,\ \ \xi^l=X^l-w^l(u)\ ,
\end{equation}
and
\begin{equation}\label{3'}
r=-\eta_{ij}\,\dot w^i\,\xi^j=-\dot w_j\,\xi^j\ .
\end{equation}

\begin{figure}[pb]
    \centerline{\includegraphics[width=0.5\textwidth]{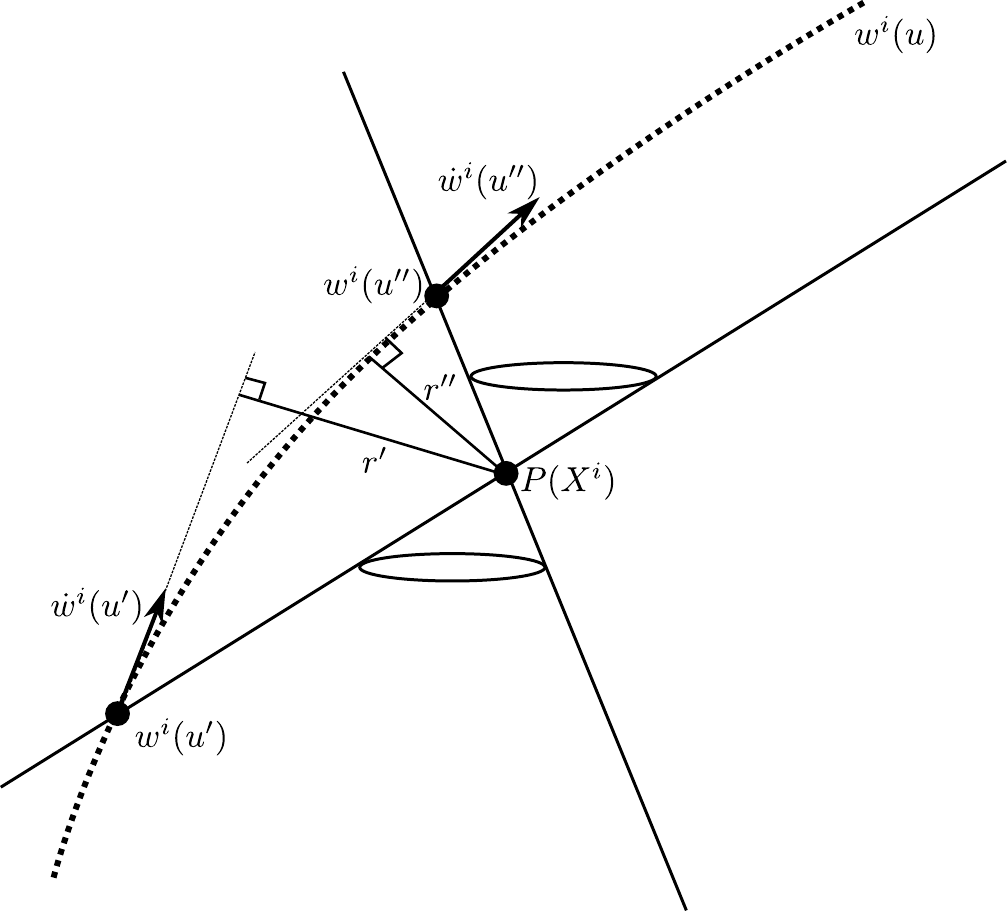}}
    \vspace*{8pt}
    \caption{\label{fig_1} The distance $r$ in eq.\ (\ref{3'}) can be the \emph{retarded distance} $r'$ of the event $P(X^i)$ to the world line $X^i=w^i(u)$ or the  \emph{advanced distance} $r''$ of $P$ to the world line \cite{Synge:1965:book}{}. For the Li\'enard--Wiechert field $r$ is the retarded distance.}
\end{figure}

\noindent
Here $\dot w^i=dw^i/du$ and in general differentiation with respect to $u$ will be indicated with a dot. Also $r$ is a measure of the distance of the event with coordinates $X^i$ from the world line, with $r=0$ if and only if $X^i=w^i(u)$, c.f.\ figure \ref{fig_1}. 

\begin{figure}[pb]
    \centerline{\includegraphics[width=0.5\textwidth]{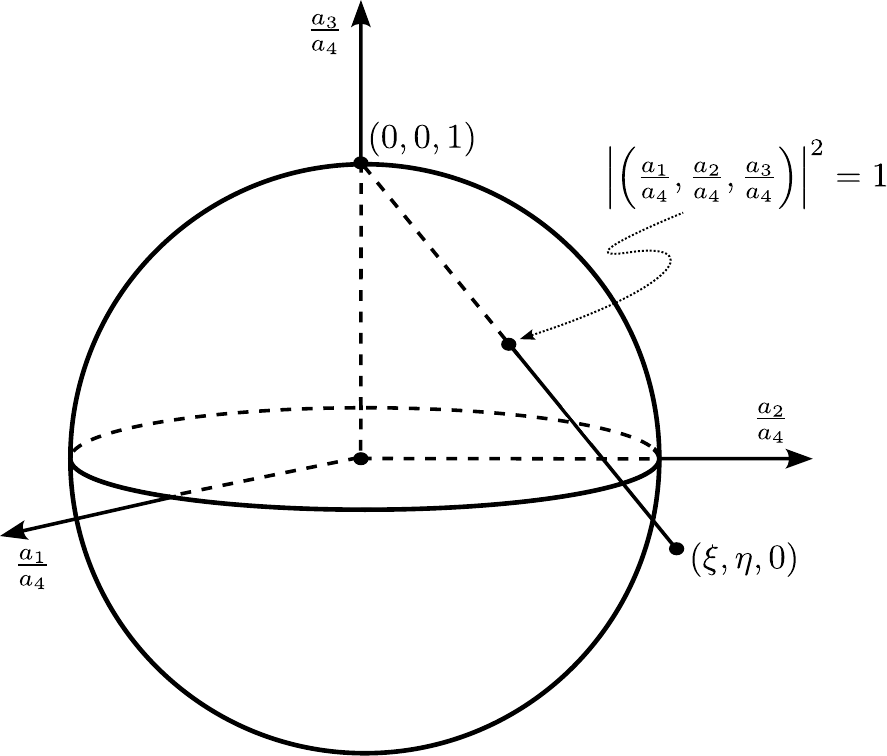}}
    \vspace*{8pt}
    \caption{\label{fig_2} In $\mathbb{R}^3$ the unit 2--sphere has equation $(a^1/a^4)^2+(a^2/a^4)^2+(a^3/a^4)^2=1$. The stereographic projection of the 2--sphere onto the equatorial plane is given by $a^1+i\,a^2=(a^4-a^3)(\xi+i\,\eta)$, $a^3=(a^4-a^3)(\xi^2+\eta^2-1)/2$ and $a^4=(a^4-a^3)(\xi^2+\eta^2+1)/2$. We are free to choose $a^4-a^3=2$ and then writing $\xi+i\,\eta=\sqrt{2}\,l(u)$ we obtain (\ref{4}).}
\end{figure}

We wish to construct a limit in which the null cones $u(X, Y, Z, T)={\rm constant}$ given by (\ref{2}) are replaced by null hyperplanes $u(X, Y, Z, T)={\rm constant}$ given by $a_i(u)\,X^i+n(u)=0$ with $\eta_{ij}\,a^i\,a^j=0$. We shall assume that $\dot n=dn/du\neq 0$ so that the null vector field $a^i(u)$ is unique up to multiplication by an arbitrary real--valued function of $u$. Thus only the \emph{direction} of this null vector field is significant as far as the equation of the null hyperplanes is concerned. This direction is determined by two real--valued functions of $u$ or equivalently by one complex--valued function $l(u)$ with complex conjugate $\bar l(u)$. How this complex--valued function of $u$ emerges is described briefly in the caption for figure \ref{fig_2}. As a result we can write the null vector field $a^i(u)$ as
\begin{equation}\label{4}
a^1+ia^2=2\,\sqrt{2}\,l(u)\ ,\ a^3+a^4=4\,l(u)\bar l(u)\ ,\ a^3-a^4=-2\ .
\end{equation}
It useful for achieving the limiting case we require (see Refs.\ \refcite{Hogan:Puetzfeld:2022:book,Hogan:Puetzfeld:2021:1}) to express the four real--valued functions $w^i(u)$ in terms of the complex--valued function $l(u)$ and two real--valued functions $m(u)$ and $n(u)$ as follows:
\begin{equation}\label{5}
w^1+iw^2=-\frac{2\,\sqrt{2}\,l}{m}\ ,\ w^3+w^4=n-\frac{4\,l\,\bar l}{m}\ ,\ w^3-w^4=\frac{2}{m}\ .
\end{equation}
Consequently
\begin{equation}\label{6}
\eta_{ij}\,\dot w^i\,\dot w^j=\frac{4\,\kappa}{m^2}\ \ {\rm with}\ \ \kappa=2\,|\beta|^2-\frac{1}{2}\alpha\,\gamma\ ,
\end{equation}
and $\beta(u)=\dot l(u)$, $\alpha(u)=\dot m(u)$, $\gamma(u)=\dot n(u)$. Thus the world line $X^i=w^i(u)$ is time--like, space--like or null depending upon whether $\kappa$ is negative, positive or zero respectively. From (\ref{3}) and (\ref{4}) we have
\begin{eqnarray}
\xi^1+i\xi^2&=&X+iY+\frac{1}{m}(a^1+ia^2)\ ,\label{7}\\
\xi^3+\xi^4&=&Z+T-n+\frac{1}{m}(a^3+a^4)\ ,\label{8}\\
\xi^3-\xi^4&=&Z-T+\frac{1}{m}(a^3-a^4)\ .\label{9}
\end{eqnarray}
Using these in (\ref{2}) we find that
\begin{equation}\label{10}
a_i\,X^i+n=\frac{n\,m}{2}(Z-T)-\frac{m}{2}\,\eta_{ij}\,X^i\,X^j\ .
\end{equation}
We see from this that if $m(u)=0$ then $u(X, Y, Z, T)$ is given implicitly by
\begin{equation}\label{11}
a_i(u)\,X^i+n(u)=0\ ,
\end{equation}
and thus the hypersurfaces $u(X, Y, Z, T)={\rm constant}$ are null hyperplanes. Clearly these null hyperplanes intersect if $\dot a^i\neq0$ ($\Leftrightarrow\ \beta(u)=\dot l(u)\neq0$ by (\ref{6})). Consequently if the null hyperplanes are the histories in Minkowskian space--time of the wave fronts of electromagnetic waves (as they turn out to be below) then these waves in general have colliding wave fronts. Next using (\ref{4}) and (\ref{6}) we have
\begin{eqnarray}
\dot w^1+i\dot w^2&=&-\frac{1}{m}(\dot a^1+i\dot a^2)+\frac{\alpha}{m^2}(a^1+ia^2)\ ,\label{11a}\\
\dot w^3+\dot w^4&=&\gamma-\frac{1}{m}(\dot a^3+\dot a^4)+\frac{\alpha}{m^2}(a^3+a^4)\ ,\label{11b}\\
\dot w^3-\dot w^4&=&\frac{\alpha}{m^2}(a^3-a^4)\ ,\label{11c}
\end{eqnarray}
and combining these with (\ref{7})--(\ref{9}) we can deduce that
\begin{equation}\label{12}
\dot w_i\,\xi^i=-\frac{1}{m}\,\dot a_i\,\xi^i+\frac{\alpha}{m^2}\,a_i\,\xi^i-\frac{\gamma}{m}\left (1-\frac{m}{2}(Z-T)\right )\ ,
\end{equation}
while
\begin{equation}\label{13}
\dot a_i\,\xi^i=\dot a_i\,X^i\ \ {\rm and}\ \ a_i\,\xi^i=a_i\,X^i+n\ ,
\end{equation}
and so, by (\ref{3'}) and (\ref{12}),
\begin{equation}\label{14}
r=\frac{1}{m}\,(\dot a_i\,X^i+\gamma)-\frac{\alpha}{m^2}\,(a_i\,X^i+n)-\frac{1}{2}\gamma\,(Z-T)\ .
\end{equation}
From this we observe that if $m(u)\rightarrow 0$ then $r\rightarrow\infty$. This will be how we will implement the asymptotic limit below when applied to a generalized Li\'enard--Wiechert electromagnetic field. 

As a final preliminary we introduce, in addition to $u$, new coordinates $\zeta, \bar\zeta$ and $v$ (following Ref.\ \refcite{Hogan:Puetzfeld:2021:1}) via
\begin{eqnarray}
X+iY&=&\sqrt{2}\,\left (1-\frac{m\,v}{2}\right )\zeta-\sqrt{2}\,l\,v\ ,\label{15}\\
Z-T&=&v\ ,\label{16}\\
Z+T&=&2\,\left (1-\frac{m\,v}{2}\right )\left\{\bar l\,\zeta+l\,\bar\zeta+\frac{1}{2}m\,\zeta\bar\zeta+\frac{1}{2}n\right\}-\left (2\,l\,\bar l-\frac{1}{2}n\,m\right )v\ ,\label{17}
\end{eqnarray}
with (\ref{17}) obtained from (\ref{10}) following substitution of (\ref{15}) and (\ref{16}) into (\ref{10}). In the coordinates $\zeta, \bar\zeta, u, v$ the Minkowskian line element 
(\ref{1}) reads
\begin{equation}\label{18}
ds^2=2\left (1-\frac{m\,v}{2}\right )^2\left |d\zeta-\frac{v\,q_{\bar\zeta}}{\left (1-\frac{m\,v}{2}\right )}du\right |^2+2\,q\,du\,dv\ ,
\end{equation}
with
\begin{equation}\label{19}
q(\zeta, \bar\zeta, u)=\bar\beta\,\zeta+\beta\,\bar\zeta+\frac{1}{2}\alpha\,\zeta\bar\zeta+\frac{1}{2}\gamma\ \ {\rm and}\ \ q_{\bar\zeta}=\frac{\partial q}{\partial\bar\zeta}\ .
\end{equation}

\section{Li\'enard--Wiechert Fields}\label{sec:3}

If $e={\rm constant}$ is the electric charge on a source having world line $X^i=w^i(u)$ in Minkowskian space--time then the potential 1--form of interest to us is
\begin{equation}\label{20}
A=\frac{e}{r}\,\dot w_i\,dX^i\ .
\end{equation}
Here the world line $X^i=w^i(u)$ ($\Leftrightarrow r=0$) is arbitrary. If the world line is time--like and $u$ is proper--time or arc length along it then (\ref{20}) is the Li\'enard--Wiechert potential 1--form. If the world line is light--like then (\ref{20}) is Synge's \cite{Synge:1972} potential 1--form. We shall initially require (\ref{20}) in the coordinates $\zeta, \bar\zeta, u, v$. With $r$ given by (\ref{14}) the transformations (\ref{15})--(\ref{17}) give us
\begin{eqnarray}
a_i\,X^i&=&-m\left (1-\frac{m\,v}{2}\right )\zeta\bar\zeta-n\ \ {\rm and}\ \nonumber \\
\dot a_i\,X^i&=&2\left (1-\frac{m\,v}{2}\right )(\bar\beta\,\zeta+\beta\,\bar\zeta)\ ,\label{21}
\end{eqnarray}
and thus (\ref{14}) becomes
\begin{equation}\label{22}
r=\frac{2\,q}{m}\left (1-\frac{m\,v}{2}\right )\ .
\end{equation}
Next with $\dot w^i$ given by (\ref{11a})--(\ref{11c}) we have
\begin{equation}\label{23}
\dot w_i\,dX^i=-\frac{1}{m}\,\dot a_i\,dX^i+\frac{1}{2}\gamma\,(dZ-dT)+\frac{\alpha}{m^2}a_i\,dX^i\ .
\end{equation}
Now the transformations (\ref{15})--(\ref{17}) yield
\begin{eqnarray}
a_i\,dX^i&=&\frac{1}{2}m^2\zeta\,\bar\zeta\,dv-m\left (1-\frac{m\,v}{2}\right )(\bar\zeta\,d\zeta+\zeta\,d\bar\zeta)\nonumber\\
&&+\left\{\frac{1}{2}m\,\alpha\,v\,\zeta\,\bar\zeta-\frac{1}{2}m\,\gamma\,v-2\left (1-\frac{m\,v}{2}\right )q\right\}du\ , \nonumber \\ \label{24}
\end{eqnarray}
and
\begin{eqnarray}
\dot a_i\,dX^i&=&-m(\bar\beta\,\zeta+\beta\,\bar\zeta)dv+2\left (1-\frac{m\,v}{2}\right )(\bar\beta\,d\zeta+\beta\,d\bar\zeta)\nonumber\\
&&-\{\alpha\,v(\bar\beta\,\zeta+\beta\,\bar\zeta)+4\,|\beta|^2v\}du\ .\label{25}
\end{eqnarray}
Consequently we can write (\ref{23}) as
\begin{eqnarray}\label{26}
\dot w_i\,dX^i&=&q\,dv-\frac{2}{m}\left (1-\frac{m\,v}{2}\right )(q_{\zeta}\,d\zeta+q_{\bar\zeta}\,d\bar\zeta) \nonumber \\
&&+\frac{2}{m}\left (\alpha\,v\,q+\kappa\,v-\alpha\,\frac{q}{m}\right )du\ .
\end{eqnarray}
Substituting (\ref{22}) and (\ref{26}) into (\ref{20}) and simplifying we find that
\begin{eqnarray}
A&=&e\left\{q^{-1}q_u+\left (1-\frac{m\,v}{2}\right )^{-1}\kappa\,q^{-1}v\right\}du \nonumber \\
&&-e\,d\left\{\log\left [m\,q\left (1-\frac{m\,v}{2}\right )\right ]\right\}\ .\label{27}
\end{eqnarray}
Here $q_u=\partial q/\partial u$ and we denote similarly partial derivatives with respect to $\zeta$ and $\bar\zeta$ below. Hence the potential 1--form of interest, modulo a pure gauge term (an exact differential in this case), reads
\begin{equation}\label{28}
A=e\left\{q^{-1}q_u+\left (1-\frac{m\,v}{2}\right )^{-1}\kappa\,q^{-1}v\right\}du\ ,
\end{equation}
and this is a 1--form field on Minkowskian space--time with line element given by (\ref{18}) and (\ref{19}). The candidate for Maxwell field is the exterior derivative of (\ref{28}) $F=dA$ given by
\begin{eqnarray}
F&=&e\left\{(\log q)_{u\zeta}-\left (1-\frac{m\,v}{2}\right )^{-1}q^{-2}q_{\zeta}\,\kappa\,v\right\}d\zeta\wedge du\nonumber \\
&&+e\Biggl\{(\log q)_{u\bar\zeta}-\left (1-\frac{m\,v}{2}\right )^{-1}q^{-2}q_{\bar\zeta}\,\kappa\,v\Biggr\}d\bar\zeta\wedge du\nonumber \\
&&-e\left (1-\frac{m\,v}{2}\right )^{-2}q^{-1}\kappa\,du\wedge dv\ .\label{29}
\end{eqnarray}
The 2--form dual to this 2--form is
\begin{eqnarray}
{}^*F&=&-i\,e(\log q)_{u\zeta}\,d\zeta\wedge du+i\,e(\log q)_{u\bar\zeta}\,d\bar\zeta\wedge du+i\,e\,q^{-2}\kappa\,d\zeta\wedge d\bar\zeta\ .\label{30}
\end{eqnarray}
From this Maxwell's equation $d{}^*F=0$ is satisfied provided
\begin{equation}\label{31}
q^2\frac{\partial^2}{\partial\zeta\partial\bar\zeta}(q^{-1}q_u)-\kappa\,q^{-1}q_u+\frac{1}{2}\kappa_u=0\ .
\end{equation}
This is automatically satisfied since
\begin{equation}\label{32}
q^2\frac{\partial^2}{\partial\zeta\partial\bar\zeta}(q^{-1}q_u)-\kappa\,q^{-1}q_u+\frac{1}{2}\kappa_u=\Psi_u-2\,q^{-1}q_u\,\Psi\ ,
\end{equation}
with
\begin{equation}\label{33}
\Psi=q^2\frac{\partial^2}{\partial\zeta\partial\bar\zeta}(\log q)+\frac{1}{2}\,\kappa\ ,
\end{equation}
and with $\kappa$ given by (\ref{5}) and $q$ by (\ref{19}) we have $\Psi=0$.

\section{Asymptotic Limit of Li\'enard--Wiechert}\label{sec:4}

The asymptotic limit of the Li\'enard--Wiechert fields in the previous section is obtained by letting $m(u)\rightarrow 0$. This results in the Minkowskian line element (\ref{18}) taking the form
\begin{equation}\label{34}
ds^2=2\,|d\zeta-\beta(u)\,v\,du|^2+2\,\hat q\,du\,dv\ \ {\rm with}\ \ \hat q=\bar\beta\,\zeta+\beta\,\bar\zeta+\frac{1}{2}\gamma\ .
\end{equation}
The $m(u)\rightarrow 0$ limit of the potential 1--form (\ref{28}) is
\begin{equation}\label{35}
A=e\,\{\hat q^{-1}\hat q_u+\kappa\,\hat q^{-1}v\}\,du\ \ {\rm with}\ \ \kappa=2\,|\beta|^2\ .
\end{equation}
To survey this vacuum electromagnetic field it is helpful to express it in coordinates $X^i=(X, Y, Z, T)$. These are obtained by taking the $m(u)\rightarrow 0$ limit of (\ref{15})--(\ref{17}) resulting in
\begin{eqnarray}
X+iY&=&\sqrt{2}\,\zeta-\sqrt{2}\,l\,v\ ,\label{36}\\
Z-T&=&v\ ,\label{37}\\
Z+T&=&2\left (\bar l\,\zeta+l\,\bar\zeta+\frac{1}{2}n\right )-2\,l\,\bar l\,v\ .\label{38}
\end{eqnarray}
Substituting for $\zeta$ and $v$ from (\ref{36}) and (\ref{37}) into (\ref{38}) we arrive at the equation giving $u(X, Y, Z, T)$ implicitly:
\begin{equation}\label{39}
a_i(u)\,X^i+n(u)=0\ ,
\end{equation}
in agreement with (\ref{11}). It thus follows that
\begin{equation}\label{40}
u_{,i}=-\frac{a_i}{\rho}\ \ {\rm with}\ \ \rho=\dot a_i\,X^i+\gamma\ .
\end{equation}
Using (\ref{36})--(\ref{38}) we find that
\begin{equation}\label{41}
\hat q=\frac{1}{2}(\dot a_i\,X^i+\gamma)=\frac{1}{2}\rho\ \ {\rm and}\ \ \hat q_u+\kappa\,v=\frac{1}{2}(\ddot a_i\,X^i+\dot\gamma)\ .
\end{equation}
As a result we can write (\ref{35}) as $A=A_i\,dX^i$ with the 4--potential having contravariant components
\begin{equation}\label{42}
A^i=-\frac{e(\ddot a_j\,X^j+\dot\gamma)}{(\dot a_k\,X^k+\gamma)^2}\,a^i\ .
\end{equation}
However
\begin{equation}\label{43}
e\,\left [\log(\dot a_j\,X^j+\gamma)\right ]_{,i}=\frac{e\,\dot a_i}{(\dot a_k\,X^k+\gamma)}-\frac{e\,(\ddot a_j\,X^j+\dot\gamma)}{(\dot a_k\,X^k+\gamma)^2}\,a_i\ ,
\end{equation}
and thus, modulo a gauge transformation, we can write the 4--potential (\ref{42}) in the simpler form
\begin{equation}\label{44}
A^i=-\frac{e\,\dot a^i}{\rho}\ .
\end{equation}
We note that 
\begin{eqnarray}
\rho_{,i}&=&\dot a_i-\rho^{-1}(\ddot a_j\,X^j+\dot\gamma)a_i\ \ \Rightarrow\ \ \eta^{ij}\,\rho_{,i}\,\rho_{,j}=\dot a_i\,\dot a^i\  {\rm and}\ \ \rho_{,i}\,a^i=0\ ,\label{45}
\end{eqnarray}
while
\begin{equation}\label{46}
\Box\rho\equiv\eta^{ij}\,\rho_{,ij}=-\frac{2}{\rho}\,\ddot a^i\,a_i=\frac{2}{\rho}\,\dot a^i\,\dot a_i\ ,
\end{equation}
and thus it straightforward to confirm that (\ref{44}) satisfies Maxwell's vacuum field equations in the form
\begin{equation}\label{47}
A^i{}_{,i}=0\ \ {\rm and}\ \ \Box A^i=0\ .
\end{equation}
The Maxwell tensor has components
\begin{equation}\label{48}
F_{ij}=A_{j,i}-A_{i,j}=\frac{\Pi_{ij}}{\rho^2}+\frac{N_{ij}}{\rho^3}\ ,
\end{equation}
with
\begin{eqnarray}
\Pi_{ij}&=&e\,(a_i\,\ddot a_j-\ddot a_i\,a_j) \quad \Rightarrow \quad \Pi_{ij}\,a^j=-e\,(\dot a_j\, \dot{a}^j)a_i\ ,\label{49}\\
N_{ij}&=&e\,(\ddot a_k\,X^k+\dot\gamma)(a_i\,\dot a_j-\dot a_i\,a_j) \quad \Rightarrow \quad N_{ij}\,a^j=0\ . \label{50}
\end{eqnarray}
We see that this asymptotic Li\'enard--Wiechert field is algebraically general with $a^i$ one of the principal null directions. In the region of Minkowskian space--time in which the dimensionless quantity $\rho=\dot a_i\,X^i+\gamma$ is small the Maxwell field is predominantly radiative with propagation direction $a^i$ and the electromagnetic radiation is plane fronted with colliding wave fronts. 

In the region of space--time corresponding to small values of $\rho$ we see from (\ref{45}) that the gradient of $\rho$ is predominantly in the direction of the propagation direction of the radiation in space--time. Also the integral curves of the gradient of $\rho$ diverge when $\rho = 0$ by (\ref{46}). This suggests that the region of space--time for which $\rho$ is small contains the history of a complicated source of the electromagnetic field. This Maxwell field was originally found by Bateman \cite{Bateman:1955:1} (his eqs.\ (299) and (302)). The significance of this solution has been mysterious up to now. The present paper elucidates its relationship to the Li\'enard--Wiechert field and this appears to be new.

Qualitatively this Bateman field is algebraically similar to the Li\'enard--Wiechert field. The Li\'enard--Wiechert field has an algebraically general part (or ``Coulomb part'') in the neighborhood in space--time of the history of the charge and an algebraically special part (or ``radiative part'') at large spatial distance from the history of the charge. The Bateman field also has two corresponding algebraic parts (\ref{48}) and (\ref{49}) which are distinguished by values of the dimensionless parameter $\rho$.

\section{Discussion}\label{sec:5}

Our motivation for the construction in sections \ref{sec:2} and \ref{sec:3} leading to the asymptotic limit of Li\'enard--Wiechert 
described in section \ref{sec:4} is the following: The Li\'enard--Wiechert 4--potential is $A^i=e\,\dot w^i(u)/r$ and its asymptotic limit, $r\rightarrow\infty$, is zero. However we found an asymptotic limit involving $r\rightarrow\infty$ accompanied by $\dot w^i(u)\rightarrow\infty$ in which the potential 1--form can have a finite, nonzero limit. Since $\dot w^i$ is a function only of $u$ we had to find a way of writing $\dot w^i$ in terms of an arbitrary function of $u$, $m(u)$ say, and a way of writing $r$ in terms of $m(u)$ so that when the asymptotic limit is triggered by $m(u)\rightarrow 0$ (say) then $A$ has a finite, nonzero limit. 

In addition, since the Li\'enard--Wiechert field contains spherical fronted electromagnetic waves we expect its asymptotic limit to contain plane fronted electromagnetic waves. This required the limit to be capable of taking us from the null cones in eq.\ (\ref{2}) to the null hyperplanes in eq.\ (\ref{11}). This involved another arbitrary function of $u$, $n(u)$ in eq.\ (\ref{11}), to avoid the null hyperplanes all intersecting at $X^i=0$. Thus implementing the asymptotic limit involved two arbitrary functions $m(u)$ and $n(u)$. This was achieved explicitly by a combination of eq.\ (\ref{22}) and eq.\ (\ref{26}) together with a gauge transformation to eventually obtain eq.\ (\ref{28}) prior to taking the limit $m\rightarrow 0$.

The original Li\'enard--Wiechert field has a region of Minkowskian space--time at large distance from the world line of the charge in which the electromagnetic field is predominantly radiative with spherical wave fronts. The asymptotic limit described here is similar in that it possesses a region of Minkowskian space--time in which the electromagnetic field is predominantly radiative but with plane wave fronts. In general the asymptotic field, in similar fashion to the Li\'enard--Wiechert field, is algebraically general with principal null directions $a^i$ and $b^i$ so that
\begin{equation}\label{51}
F_{ij}=e\,\frac{k^2}{\rho^2}(a_i\,b_j-a_j\,b_i)\ ,
\end{equation}
with $k^2=\dot a_i\,\dot a^i$ and 
\begin{eqnarray}
b_i&=&k^{-2}\ddot a_i+\rho^{-1}k^{-2}(\ddot a_j\,X^j+\dot\gamma)\dot{a}_i \nonumber\\
&&+\Biggl\{\frac{1}{2}k^{-4}\ddot a_j\,\ddot a^j+\frac{1}{2}k^{-2}\rho^{-2}(\ddot a_j\,X^j+\dot\gamma)^2+k^{-3}\dot k\,\rho^{-1}(\ddot a_j\,X^j+\dot\gamma)\Biggr\}a_i\ .  \label{52}
\end{eqnarray}
Thus $b^i\,b_i=0$ and $b^i\,a_i=-1$.

It would be interesting to discover the gravitational analogue of the Maxwell field described in this paper.

\bibliographystyle{ws-ijmpd}
\bibliography{planeem_bibliography}
\end{document}